\documentclass[prb,twocolumn,aps,showpacs,floatfix]{revtex4}
\usepackage[english]{babel}
\usepackage{epsfig}
\usepackage{graphicx}
\usepackage{subfigure}
\usepackage{amsmath,amssymb,amsfonts}
\usepackage{bm}

\usepackage{color}
\usepackage{xcolor}
\usepackage{ulem}

\begin{document}

\title{Relationship between orbital structure and lattice distortions: CE phase of manganites, revisited}

\author{A.\,O.~Sboychakov, K.\,I.~Kugel, and A.\,L.~Rakhmanov}
\affiliation{Institute for Theoretical and Applied
Electrodynamics, Russian Academy of Sciences, Izhorskaya Str. 13, Moscow, 125412 Russia}

\author{D.~I.~Khomskii}
\affiliation{$II.$ Physikalisches Institut, Universit\"at zu
K\"oln, Z\"ulpicher Str.~77, 50937 K\"oln, Germany}

\begin{abstract}

Analyzing the orbital structure and lattice distortions in the CE phase of half-doped manganites, we demonstrate that the usual approach directly relating the orbital occupation of Jahn-Teller ions to the displacements of neighboring ligands may be misleading. For the correct identification of orbital structure, it is necessary to take self-consistently into account the electron-lattice interactions, kinetic energy of charge carriers, and crystal-field effects. In certain situations, e.g. in the CE phase of single-layered manganite La$_{0.5}$Sr$_{1.5}$MnO$_4$, the type of orbital ordering strongly deviates from that, which one would deduce from the local lattice distortions.

\end{abstract}



\date{\today}

\maketitle

\section{Introduction}

An interplay between spin, charge, orbital, and lattice degrees of freedom play an important role in the physics of transition-metal oxides, especially those with Jahn-Teller (JT) ions, such as Mn$^{3+}$ or Cu$^{2+}$. In the materials with JT ions an orbital ordering  and related lattice distortions determine a rich variety of different phenomena, and this ``orbital physics" attracts now a widespread attention~\cite{TokNagSci,dagbook,vdBrKhaKhoRev}. Unfortunately, it is rather difficult to directly access orbital state of JT ions. Therefore, the standard way to find orbital occupation, widely used for the last 50 years, is based on structural data, with the assumption of the one-to-one correspondence between orbital occupation of an ion and corresponding JT distortion of the surrounding ligands, e.g. O$_6$ octahedra~\cite{Gooden}. However, as we show below, this straightforward approach sometimes fails for the solids with orbital ordering. A  well-studied layered manganite La$_{0.5}$Sr$_{1.5}$MnO$_4$ gives a vivid example of such discrepancy. Below 240 K, this compound exhibits a charge ordering characterized by the  checkerboard arrangement of the Mn$^{3+}$ and Mn$^{4+}$ ions. Below $T_N$ =110 K, a spin ordering of the CE type appears consisting of zigzag ferromagnetic (FM) chains with an antiferromagnetic (AFM) stacking between them. The CE-type AF was ascribed to an orbital order at the formal Mn$^{3+}$ sites~\cite{GoodenPR55,WolKoeh}. At the same time, the detailed type of orbital order involving a spatial redistribution of valence electrons is still controversially discussed in literature. Some experiments are more compatible with the $3x^2-r^2/3y^2-r^2$ order~\cite{Mirone,HuaWu} originally suggested in Refs.~\onlinecite{GoodenPR55,WolKoeh}, while the other suggest the  $x^2-z^2/y^2-z^2$ order~\cite{Wilk1,Wilk2,Huang}.

The theoretical treatment based only on electronic mechanism  of orbital ordering ~\cite{vdBrKhaKho,Dong} gives a rather good description of half-doped manganites, which reproduces the original assignment of occupied $3x^2-r^2/3y^2-r^2$ orbitals. However, to get a better insight into an actual physical situation, one should also include explicitly electron-lattice interaction as well as crystal-field effects. For undoped manganites, an attempt in this direction was recently undertaken in Ref.~\onlinecite{Yarlag}. In the present paper, we analyze the interrelation of orbital order, lattice distortions, and charge disproportionalization for FM zigzag chains in CE phase of half-doped manganites, and demonstrate that the standard paradigm of one-to-one correspondence of orbital ordering and distortion of nearest neighbor ligands fails in these cases. Our results may be a ``warning sign" that similar discrepancy may also occur in other situations, and that the standard way to obtain orbital occupation from the local lattice distortions may be inapplicable in some cases.

\section{A model of zigzag chains in half-doped manganites}

\subsection{The model Hamiltonian}

\begin{figure}[hb]
\begin{center}
\includegraphics*[width=0.95\columnwidth]{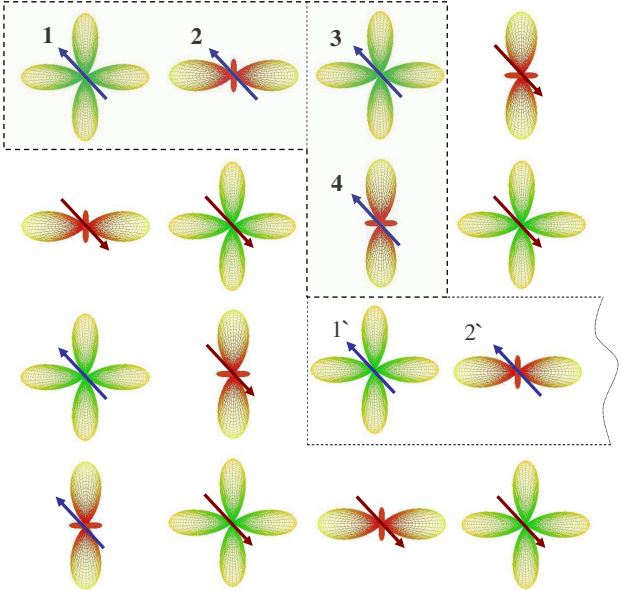}
\end{center}
\caption{\label{FigCE} (Color online) Orbital and magnetic structures of Mn ions in one of MnO$_2$ planes of the CE phase, originally proposed in Refs.~\onlinecite{GoodenPR55,WolKoeh}.
A ``building block" of the zigzag chain consisting of four sites is marked by a dashed line.
}
\end{figure}

Let us consider an orbital structure in CE phase of manganites. In CE phase, the system consists of zigzag ferromagnetic (FM) chains, and the neighboring chains are ordered antiferromagnetically, as it is shown in Fig.~\ref{FigCE}.

This structure can occur e.g. in layered manganites such as La$_{2-x}$Sr$_{x}$MnO$_4$ when $x\approx 1.5$, or perovskite manganites such as La$_{1-x'}$Ca$_{x'}$MnO$_3$ near half doping, when $x'\approx 0.5$. In both these cases, a charge ordering (CO) occurs with decreasing temperature: we have the checkerboard CO in the basal plane of, formally, Mn$^{3+}$ ($t_2g^3e_g^1$) and Mn$^{4+}$ ($t_2g^3e_g^0$) ions, with Mn$^{3+}$ being strong Jahn-Teller ions (one electron on a doubly degenerate $e_g$  orbital). in these systems, there occurs an orbital ordering (OO) simultaneously with the charge ordering, and the OO originally proposed in Refs.~\onlinecite{GoodenPR55,WolKoeh} is shown in Fig.~\ref{FigCE}. As mentioned in the Introduction, it is just this detailed type of OO, which is now put under question~\cite{Huang}, and this is the main problem we address theoretically in our work.

The model Hamiltonian describing this situation can be written as
\begin{eqnarray}\label{HCE}
H =-\sum_{\langle{nm}\rangle\alpha\beta}\left(t^{nm}_{\alpha\beta}
{a}^{\dag}_{n\alpha}{a}_{m\beta}+h.c.\right)%
+g\sum_{n}\left(Q_{2n}{\tau}^x_{n}+Q_{3n}{\tau}^z_{n}\right) \nonumber \\
+\sum_{n}\left(K\frac{Q_{2n}^2+Q_{3n}^2}{2}
-\sum_n\varepsilon_zQ_{3n}\right)
- \sum_n\Delta\tau^z_n
\end{eqnarray}

\begin{figure}
\begin{center}
\includegraphics*[height=0.15\textheight]{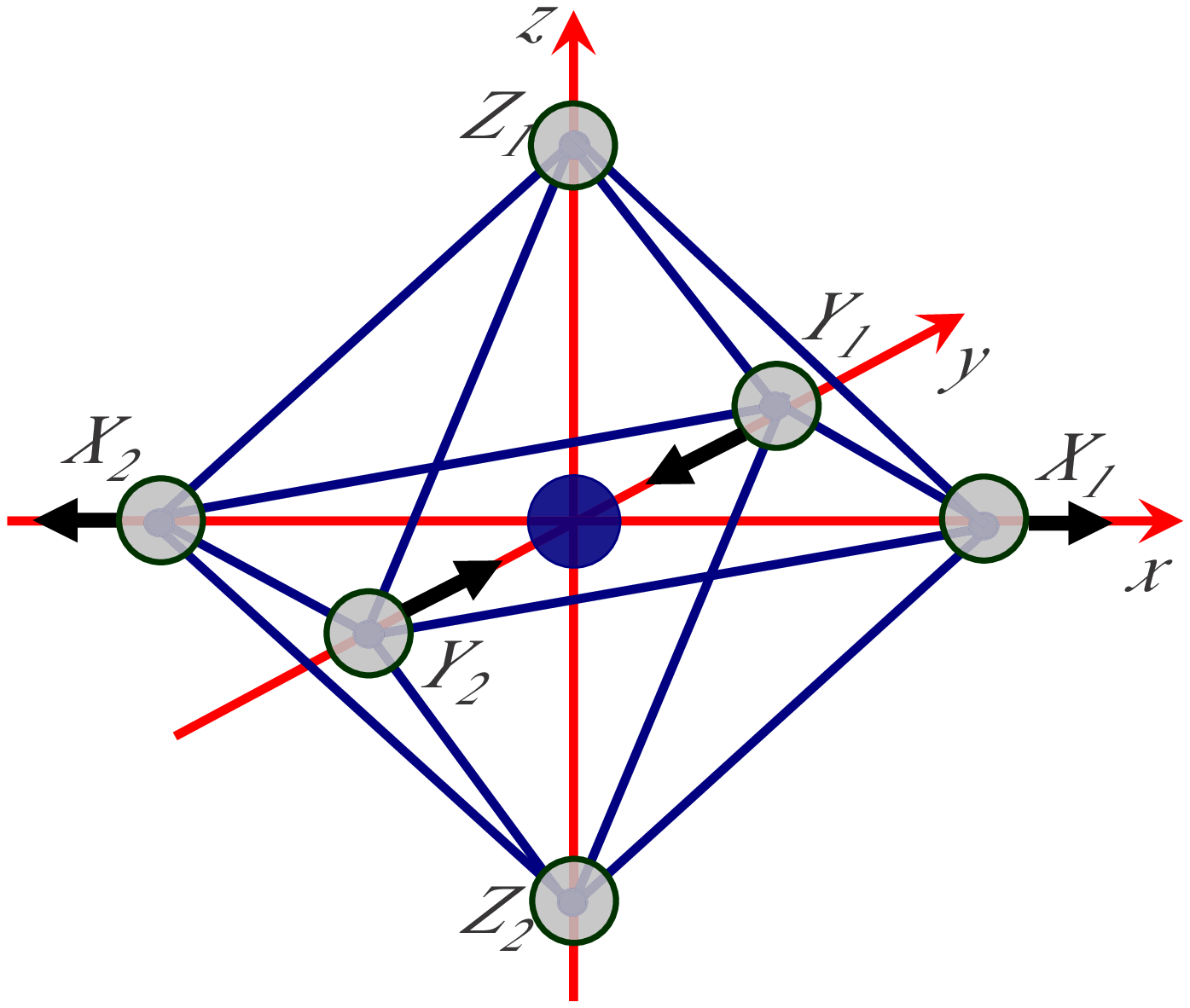}\hspace{1cm}
\includegraphics*[height=0.15\textheight]{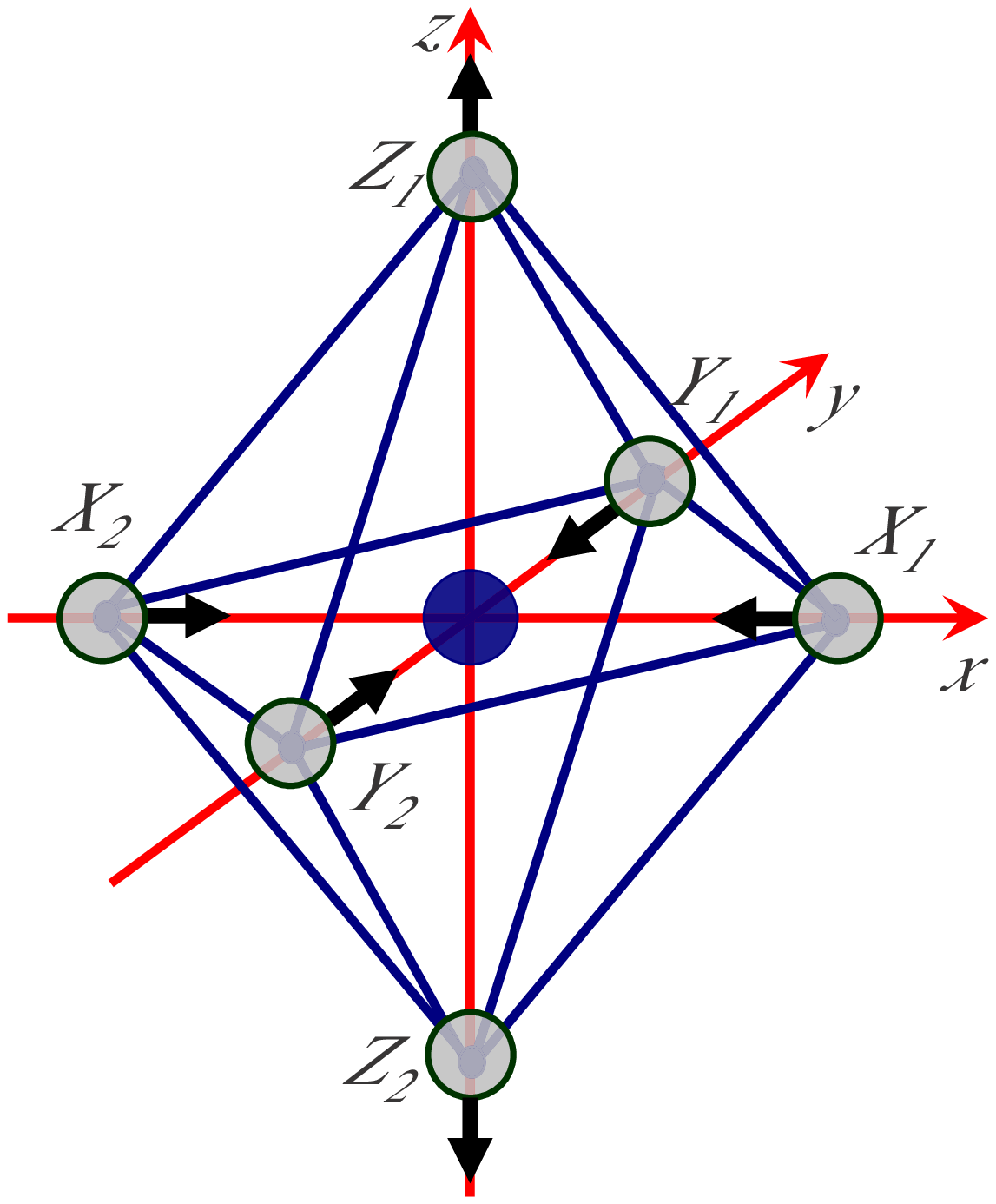}
\end{center}
\caption{\label{FigQs} (Color online) $Q_2$--type (left panel) and $Q_3$--type (right panel) distortions of a MnO$_6$ octahedron.}
\end{figure}

In this expression, ${a}^{\dag}_{n\alpha}$ and ${a}_{n\alpha}$ are creation and annihilation operators  for an $e_g$ electron in orbital state $\alpha$ ($x^2-y^2$ or $2z^2-x^2-y^2$) at a site $n$, $Q_{2n}$ and $Q_{3n}$ describe Jahn-Teller (JT) distortions of MnO$_6$ octahedron (see, Fig.~\ref{FigQs}), $\varepsilon_{z}$ is the $zz$ component of the tensor, and ${\tau}^{x,z}_{n}$ are pseudospin operators
\begin{equation}\label{tau}
{\tau}^{x,y,z}_{n}=\sum_{\alpha\beta}{a}^{\dag}_{n\alpha}
\sigma^{x,y,z}_{\alpha\beta}{a}_{n\beta}\,,
\end{equation}
describing orbital occupation of $e_g$ orbitals; here $\sigma^{x,y,z}_{\alpha\beta}$ are Pauli matrices. Note, that in our model the spin of $e_g$ electrons is always parallel to the core spin $S$ of $t_{2g}$ electrons. Moreover, all spins in the chain are assumed to be ferromagnetically ordered, and we omit spin indices in Hamiltonian~\eqref{HCE}. We consider below one zigzag chain: as is common in such cases~\cite{DeGennes} for the CE magnetic ordering of Fig.~\ref{FigCE} there will be no electron hopping between zigzags with antiparallel spins, and the elastic coupling between zigzags, which may be present, does not modify the results.

The first term in Eq.~\eqref{HCE} is the kinetic energy of $e_g$ electrons. The hopping amplitudes depend both on the positions of neighboring sites $n$ and $m$, and on the orbital states of the $e_g$ electron before and after hopping. They can be written as
\begin{equation}\label{txy}
t^{nm}_{\alpha\beta}=\frac{t_0}{4}\left(%
\begin{array}{cc}
3 & \mp\sqrt{3} \\
\mp\sqrt{3} & 1 \\
\end{array}%
\right),
\end{equation}
where minus (plus) sign corresponds to the hopping in $x$ ($y$) direction, see Fig.~\ref{FigCE}. From the relation between $t^{nm}_{\alpha\beta}$ for different $\alpha$ and $\beta$, it can be found that the kinetic energy term favors the $2x^2-y^2-z^2$ and $2y^2-z^2-x^2$ orbital states in the bridge sites of the CE chain (sites $2$ and $4$ in Fig.~\ref{FigCE}), while at corner sites (sites $1$ and $3$ in Fig.~\ref{FigCE}) the $x^2-y^2$ states  turn out to be more favorable~\cite{vdBrKhaKho}.

The second term in Eq.~\eqref{HCE} is the electron-lattice interaction, while the third term is the elastic energy. The $Q_{3n}$ mode describes the stretching if $Q_{3n}>0$  and compression in the opposite case) of MnO$_6$ octahedron in $z$ direction, alongside with the compression in $x$ and $y$ directions, while $Q_{2n}$ mode describes the stretching if $Q_{2n}>0$ in $x$ direction (and compression in $y$ direction of MnO$_6$ octahedron (see Fig.~\ref{FigQs}). In both cases, the volume of the octahedron does not change. Note that we neglect the breathing mode $Q_{1n}$ in Eq.~\eqref{HCE}, describing the uniform stretching or compression.

In addition to the usual terms describing the electron hopping and JT interaction, we introduced in Hamiltonian~\eqref{HCE} two extra terms, the 4th and the 5th terms in Eq.~\eqref{HCE}. $\varepsilon_z$ describes an external stress acting on MnO$_6$ octahedra, which tends to elongate them in $z$ direction (for $\varepsilon_z>0$) or compress them for $\varepsilon_z <0$. Such term may appear just due to crystal structure, or e.g. for films it can describe the strain caused by the lattice mismatch with the substrate. Specifically, in the layered manganites like La$_{2-x}$Sr$_x$MO$_4$ there exists the tendency to local elongation of MnO$_6$ octahedra in $z$ direction (i.e. $\varepsilon_z>0$), determined just by the crystal structure and leading to such elongation even in the absence of orbital degeneracy, e.g. for M=Ni$^{2+}$ in La$_2$NiO$_4$ (see Ref.~\onlinecite{La2NiO4}). The Jahn-Teller effect for ions M with orbital degeneracy will operate on the background of this strain due to a layered structure.

The last term in Hamiltonian~\eqref{HCE} describes an eventual contribution to crystal field splitting due to further nearest neighbors. Again, for layered materials like A$_2$MO$_4$ the interaction with further neighbors reduces local symmetry from cubic to (at least) tetragonal, which would lead to a crystal-field splitting of $e_g$ states even for regular MnO$_6$ octahedra. In what follows, we will separately study the possible role of strain effects ($\varepsilon_z$ terms in Hamiltonian~\eqref{HCE}) and the longer-range contribution to crystal field ($\Delta$-term in Eq. ~\eqref{HCE}) on the orbital occupation vs lattice distortion.

\subsection{Fully localized case}

First, let us find the orbital structure of the zigzag chain in the absence of electron hopping between lattice sites. A particular orbital occupation for $e_g$ electron at site $n$  is caused  by the splitting of the doubly degenerate $e_g$ levels, connected in particular with the JT distortions, $Q_{2n}$ and  $Q_{3n}$. The values of $Q_{2n}$ and  $Q_{3n}$ depend on the orbital state of $e_g$ electron at site $n$. If we neglect the intersite hopping, these distortions can be found analytically. In this case, the $e_g$ electron at a site $n$ can be described by the wave function $|\theta_n\rangle$, which is the superposition of basic states, $|a\rangle=|x^2-y^2\rangle$ and $|b\rangle=|2z^2-x^2-y^2\rangle$
\begin{equation}\label{theta}
|\theta_n\rangle=-\sin\frac{\theta_n}{2}|a\rangle+
\cos\frac{\theta_n}{2}|b\rangle\,.
\end{equation}

Similarly, one can introduce an angle $\theta'$ characterizing local distortion of MO$_6$ octahedra in ($Q_2$, $Q_3$) plane
\begin{equation}\label{theta'}
\tan\theta'_n= \frac{Q_{2,n}}{Q_{3,n}}\,.
\end{equation}

In the usual approach, one always assumes that the orbital mixing angle $\theta$ is equal to the distortion mixing angle $\theta'$, $\theta = \theta'$. If so, one could indeed obtain orbital occupation from the measure of local M-O distances, as is usually done. We will show below that, generally speaking, these mixing angles may be different, $\theta \neq \theta'$, i.e. the orbital ordering does not necessarily follows local distortion.

The energy corresponding to the state $|\theta_n\rangle $ is found from Eq.~\eqref{HCE} (with $t^{{\bf nm}}_{\alpha\beta}=0$) using obvious relations $\tau^{z}_n|\theta_n\rangle =|-\theta_n\rangle$ and $\tau^{x}_n|\theta_n\rangle =|\pi-\theta_n\rangle$. Minimizing the total energy with respect to $Q_{2n}$ and $Q_{3n}$, we find
\begin{equation}\label{Qtheta}
{Q}^{\theta}_2=\frac{g\sin\theta_n}{K},
\;\;\;{Q}^{\theta}_3=\frac{g\cos\theta_n+\varepsilon_z}{K}\,.
\end{equation}

Already from here, we see that for $\varepsilon_z \neq 0$ local distortions $Q_2$ and $Q_3$ do not exactly follow the orbital occupation (orbital mixing angle is $\theta_n$, but $Q_2/Q_3 \neq \tan\theta_n$, as it would if the angles $\theta_n$ and $\theta'_n$ would be the same).

The total energy then reads
\begin{equation}\label{EQtheta}
\bar{E}^{\theta}=\sum_n\left[-\frac{g^2+\varepsilon_z^2}{2K}-
\tilde{\Delta}\cos\theta_n\right]\,,\;\;\;\tilde{\Delta} = \frac{g\varepsilon_z}{K}+\Delta \, .
\end{equation}

When $\tilde{\Delta}=0$, $\bar{E}^{\theta}$ does not depend on $\theta_n$, and all states $|\theta_n\rangle$ are equivalent. The non-zero $\tilde{\Delta}$ breaks this symmetry, making the state $|2z^2-x^2-y^2\rangle$ ($\theta_n=0$) when $\tilde{\Delta}>0$, or the state $|x^2-y^2\rangle$ ($\theta_n=\pi$) when $\tilde{\Delta}<0$, more favorable in energy. In addition, the non-zero stress $\varepsilon_z$ leads to non-zero distortion $Q_{3n}$ even without $e_g$ electrons. In equilibrium, this distortion is $\bar{Q}_{3}^{(0)}=\varepsilon_z/K$. The non-zero $\bar{Q}_{3}^{(0)}$ leads to the additional splitting of the $e_g$ levels with the splitting energy $\Delta_{0}=2g\bar{Q}_{3}^{(0)}=2g\varepsilon_z/K$.

\subsection{Non-zero hopping: the mean-field approximation}

For non-zero hopping, the wave functions $|\theta_n\rangle$ are not eigenfunctions of the  Hamiltonian~\eqref{HCE}, and it is necessary to diagonalize ${H}$ numerically or using some approximations. In a mean field (MF) approximation, the JT distortions are found by minimization of Hamiltonian~\eqref{HCE} averaged over electronic degrees of freedom. It gives
\begin{equation}\label{Q}
\bar{Q}_{2n}=-\frac{g\langle\tau^x_{n}\rangle}{K}\,,\;\;\;%
\bar{Q}_{3n}=-\frac{g\langle\tau^z_{n}\rangle-\varepsilon_z}{K}\,.
\end{equation}

This expression is a generalization of Eq.~\eqref{Qtheta} to the case of nonzero intersite hopping. As we will show below, the orbital occupation characterized by average pseudospins $\langle \tau^z\rangle$ and $\langle \tau^x \rangle$ is determined not only by the local distortion  of the first coordination sphere (here, O$_6$ octahedron) but also by other factors such as a contribution of further neighbors to the crystal-field splitting, or by the effect of band formation. However, given the orbital occupation (values of $\langle \tau^z\rangle$ and $\langle \tau^x \rangle$), we can find local distortions (the values of $\bar{Q}_2$ and $\bar{Q}_3$) from the expression~\eqref{Q}, i.e. these local distortions are determined by the orbital occupation and by the ``external" stress $\varepsilon_z$.

Substituting Eq.~\eqref{Q} into Hamiltonian~\eqref{HCE}, we obtain an effective electron Hamiltonian, which is quadratic in electronic operators, and which can be easily diagonalized. The mean values $\langle\tau^x_{n}\rangle$ and $\langle\tau^z_{n}\rangle$ are found using this effective Hamiltonian, and as a result, we find a self-consistent equations for $\langle\tau^{x,z}_{n}\rangle$ and $\bar{Q}_{2,3n}$.

In this procedure, we assume the existence of some superstructure in the system. In the case of CE phase, we have a one-dimensional effective electronic Hamiltonian. The unit cell of the zigzag FM chain consists of 4 Mn ions (we enumerate them by Latin subscripts $i,\,j,\ldots=1,\,2,\,3,\,4$, see Fig.~\ref{FigCE}), whereas $\langle\tau^{x,z}_{n}\rangle$ and $\bar{Q}_{2,3n}$ are equal for sites in equivalent positions. However, due to internal symmetry of CE chain, the number of sites in the unit cell can be reduced to two. Indeed, sites $3$ and $4$ can be made equivalent to sites $1$ and $2$, respectively, through the use of a certain transformation. Note that the JT distortions, $\bar{Q}_{2i}$ and $\bar{Q}_{2i}$, obey the following relationships
\begin{eqnarray}\label{Qsymm}
\bar{Q}_{23}=-\bar{Q}_{21},\;\;\bar{Q}_{33}=+\bar{Q}_{31},\nonumber \\
\bar{Q}_{24}=-\bar{Q}_{22},\;\;\bar{Q}_{34}=+\bar{Q}_{32}\,.
\end{eqnarray}
Then, we introduce new electron operators $c_{nj\alpha}$ according to the following formulas:
\begin{equation}\label{trans1}
\left(\begin{array}{c}c_{nja}\\ c_{njb}\end{array}\right)=%
\left(\begin{array}{cc}
e^{\mp\frac{i\pi}{2}}&0\\0&1
\end{array}\right)\left(\begin{array}{c}a_{n_{j}a}\\
a_{n_{j}b}\end{array}\right),\;\;\;j=1,\,3\,,
\end{equation}

\begin{equation}\label{trans2}
\left(\begin{array}{c}c_{nja}\\ c_{njb}\end{array}\right)=%
\left(\begin{array}{cc}
e^{\mp\frac{i\pi}{2}}&0\\0&1
\end{array}\right)
\left(\begin{array}{cc}
\frac12&\pm\frac{\sqrt{3}}{2}\\ \mp\frac{\sqrt{3}}{2}&\frac12
\end{array}\right)
\left(\begin{array}{c}a_{n_{j}a}\\a_{n_{j}b}\end{array}\right),\;\;\;j=2,\,4\,,
\end{equation}
where $n_{j}$ enumerate sites equivalent to the site $j$, and the upper (lower) sign corresponds to $j=1,\,2$ ($j=3,\,4$). The transformation~\eqref{trans1} for corner sites ($j=1,\,3$) is reduced to the $\mp\pi/2$ phase shift in $a_{n_ja}$ operators, while the transformation~\eqref{trans2} for bridge sites ($j=2,\,4$) consists of two subsequent transformations: the rotation in the pseudospin space by an angle $\theta=2\pi/3$ for $j=2$ and $\theta=-2\pi/3$ for $j=4$, and the additional phase shift. Note, that the rotation by an angle $\theta=2\pi/3$ ($\theta=-2\pi/3$) corresponds to choosing the states $|y^2-z^2\rangle$ and $|2x^2-y^2-z^2\rangle$  ($|z^2-x^2\rangle$ and $|2y^2-x^2-z^2\rangle$ ) as new basic states. After these transformations, sites $3$ and $4$ become equivalent to sites $1$ and $2$, respectively.

Thus, we can introduce a new unit cell consisting of two Mn ions, and a new electron creation and annihilation operators, $c^{\dag}_{nA}$ and $c_{nA}$, acting on the states inside the two-site unit cell. They can be represented in the form of a vector
\begin{equation}\label{cA}
c^{\dag}_{nA}=\left(\begin{array}{cccc}c^{\dag}_{n1a}&c^{\dag}_{n1b}&c^{\dag}_{n2a}&
c^{\dag}_{n2b}\end{array}\right),\;\;
c_{nA}=\left(\begin{array}{c}c_{n1a}\\c_{n1b}\\c_{n2a}\\
c_{n2b}\end{array}\right),
\end{equation}
where $n$ enumerates now new unit cells. Using relationships~\eqref{trans1}, \eqref{trans2}, and Eq.~\eqref{tau}, we can rewrite Hamiltonian~\eqref{HCE} in terms of the new electron operators. In the momentum representation the effective Hamiltonian can be written as
\begin{equation}\label{Heff}
{H}_{\text{eff}}=\sum_{kAB}c^{\dag}_{kA}{\varepsilon}^{AB}(k)c_{kB}+%
\sum_{n}\left(K\frac{\bar{Q}_{2n}^2+\bar{Q}_{3n}^2}{2}-
\varepsilon_z\bar{Q}_{3n}\right),
\end{equation}
where
\begin{widetext}
\begin{equation}\label{es}
\hat{\varepsilon}(k)=\left(\begin{array}{cccc}g\tilde{Q}_{31}
&g\bar{Q}_{21}&0&\displaystyle\frac{i}{2}t_0\sqrt{3}\left(1-e^{-ik}\right)\\%
g\bar{Q}_{21}&-g\tilde{Q}_{31}&0&-\displaystyle\frac12t_0\left(1+e^{-ik}\right)\\%
0&0&\displaystyle\frac{g}{2}\left(\bar{Q}_{22}\sqrt{3}-
\tilde{Q}_{32}\right)&-\displaystyle\frac{g}{2}
\left(\bar{Q}_{22}+\tilde{Q}_{32}\sqrt{3}\right)%
\\-\displaystyle\frac{i}{2}t_0\sqrt{3}\left(1-e^{ik}\right)&-
\displaystyle\frac12t_0\left(1+e^{ik}\right)
&-\displaystyle\frac{g}{2}\left(\bar{Q}_{22}+\tilde{Q}_{32}\sqrt{3}\right)&%
-\displaystyle\frac{g}{2}\left(\bar{Q}_{22}\sqrt{3}-
\tilde{Q}_{32}\right)\end{array}\right),
\end{equation}
\end{widetext}
and $\tilde{Q}_{3i} = \bar{Q}_{3i}-\Delta/g$.

\begin{figure}
\begin{center}
\includegraphics[width=0.95\columnwidth]{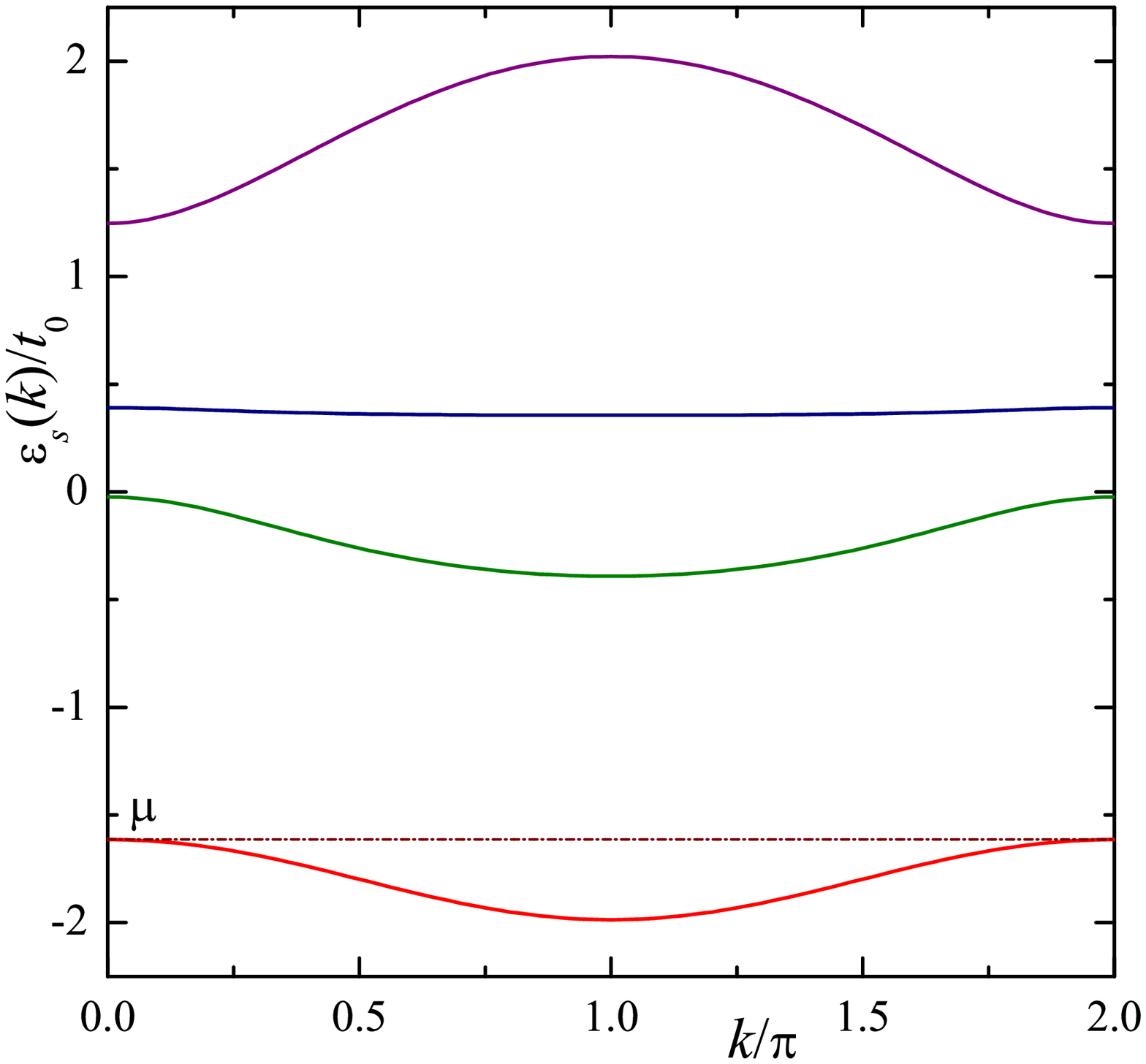}\\
\includegraphics[width=0.95\columnwidth]{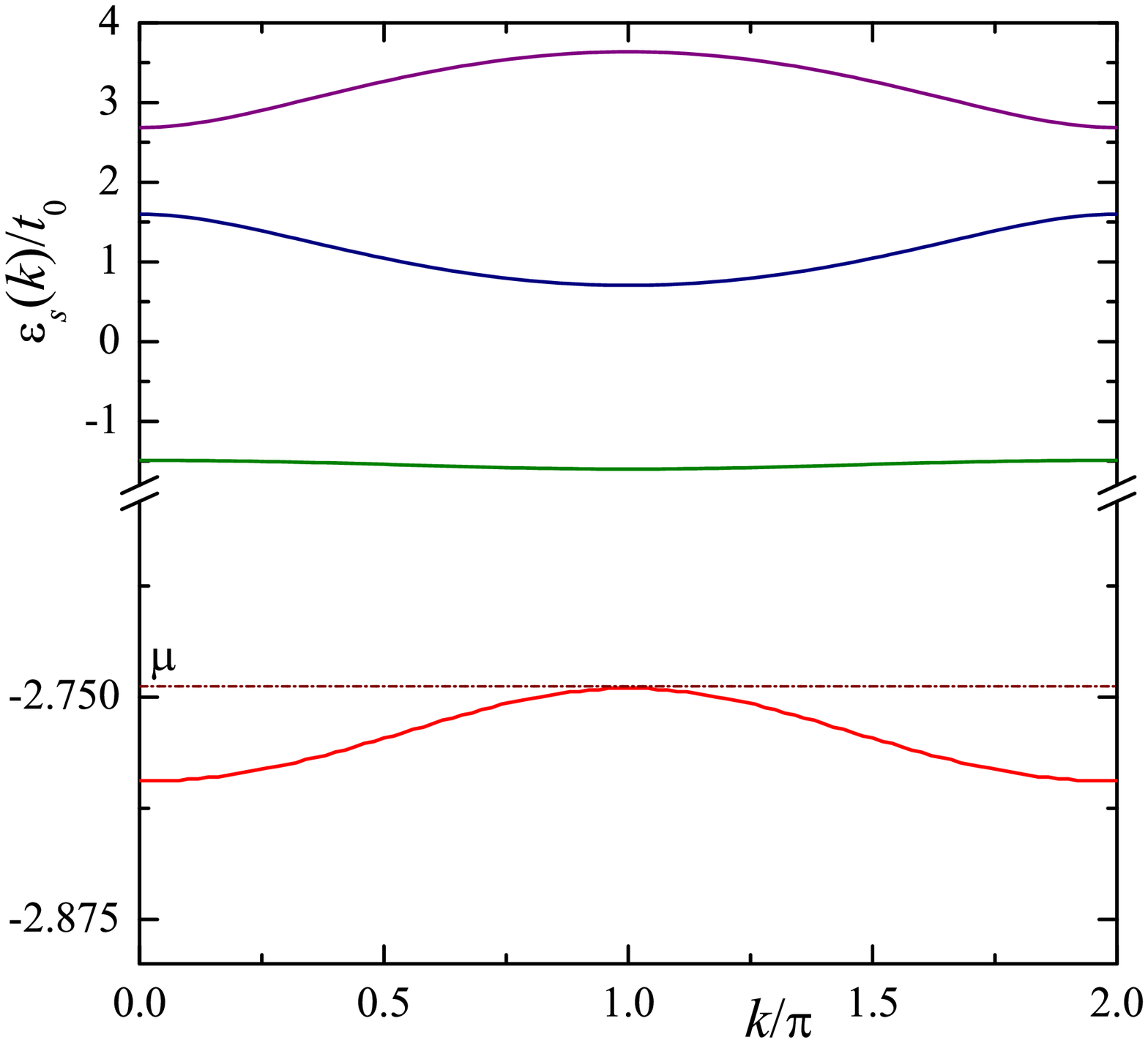}
\end{center}
\caption{\label{FigBand} (Color online) The spectrum of electrons in CE chain $\varepsilon_s(k)$  ($s=1,\,2,\,3,\,4$ from bottom to top) at small ($\varepsilon_z= 0.4$, upper panel) and large ($\varepsilon_z =1.5$, lower panel) values of the stress. $\varepsilon_z$ is measured in units of $Kt_0/g$. The parameters of the model are: $g^2/Kt_0=1.2$, $K/g=40$, $\Delta = 0$. $\mu$ is the chemical potential. In the lower panel, the lowest band $s=1$ is shown in a larger scale.}
\end{figure}

\begin{figure}
\begin{center}
\includegraphics[width=0.95\columnwidth]{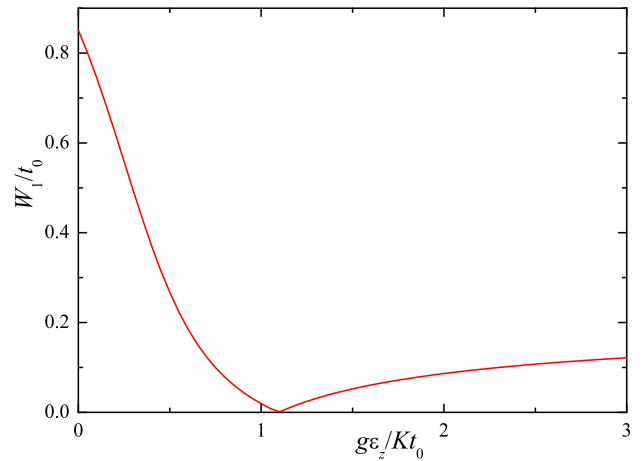}
\end{center}
 \caption{\label{FigBandWidth} (Color online) The width $W_1$ of the lowest band ($s=1$) as function of strain. The parameters of the model are: $g^2/Kt_0=1.2$, $K/g=40$.}
\end{figure}

The spectrum of electrons in CE chain is found by diagonalization of this matrix. It consists of four non-intersecting bands separated from each other by energy gaps. In the case of half-filling, $x=0.5$, there is one spinless electron for two sites. Since the bands do not overlap, only the lowest band is filled at $x=0.5$. The spectra of electrons at different values of stress $\varepsilon_z$ are shown in Fig.~\ref{FigBand}. At small $\varepsilon_z$, the spectrum of the lowest band has a minimum at $k =\pi$, while at large $\varepsilon_z$, $k =\pi$ corresponds to the maximum. Hence at some value of $\varepsilon_z$, the spectrum flattens resulting in zero bandwidth. The evolution of width $W_1$ of the lowest band with the growth of stress is illustrated in Fig.~\ref{FigBandWidth}.

Let us denote the eigenvalues of $\hat{\varepsilon}(k)$ by $\varepsilon_s(k)$ and its eigenvectors by $v_A^{(s)}(k)$ ($s=1,\,2,\,3,\,4$). The mean values $\langle c^{\dag}_{nA}c_{nB}\rangle$ can be written as
\begin{equation}\label{meanC}
\langle c^{\dag}_{nA}c_{nB}\rangle=\sum_{s}\int\limits_{0}^{2\pi}
\frac{dk}{2\pi}v_B^{(s)}(k)v_A^{*(s)}(k)\theta(\mu-\varepsilon_s(k))\,,
\end{equation}
where chemical potential $\mu$ is found from the condition
\begin{equation}
n=1-x=\frac14\sum_A\langle c^{\dag}_{nA}c_{nA}\rangle\,.
\end{equation}

In terms of new electron operators, expressions~\eqref{Q} for the mean values of JT distortions can be rewritten as
\begin{equation}\label{Q1}
\bar{Q}_{21}=\frac{g\langle\tilde{\tau}^y_{n1}\rangle}{K}\,,\;\;\;%
\bar{Q}_{31}=-\frac{g\langle\tilde{\tau}^z_{n1}
\rangle-\varepsilon_z}{K}\,,
\end{equation}
\begin{eqnarray}\label{Q2}
\bar{Q}_{22}&=&-\frac{g\left[\langle\tilde{\tau}^z_{n2}\rangle\sqrt{3}
+\langle\tilde{\tau}^y_{n2}\rangle\right]}{2K} \nonumber \\
\bar{Q}_{32}&=&-\frac{g\left[-\langle\tilde{\tau}^z_{n2}\rangle
+\langle\tilde{\tau}^y_{n2}\rangle\sqrt{3}\right]-2\varepsilon_z}{2K}\,,\;\;\;%
\end{eqnarray}
where
\begin{equation}\label{tauC}
\tilde{\tau}^{x,y,z}_{ni}=\sum_{\alpha\beta}
{c}^{\dag}_{ni\alpha}\sigma^{x,y,z}_{\alpha\beta}{c}_{ni\beta}\,.
\end{equation}
Eqs.~\eqref{es}-\eqref{Q2} together with relationships~\eqref{Qsymm} form a closed system of equations for finding both JT distortions and electronic configurations at corner and bridge sites of the CE chain.

\section{Results}

\subsection{Orbital states and local lattice distortions at bridge sites}

\begin{figure}
\begin{center}
\includegraphics[width=0.45\textwidth]{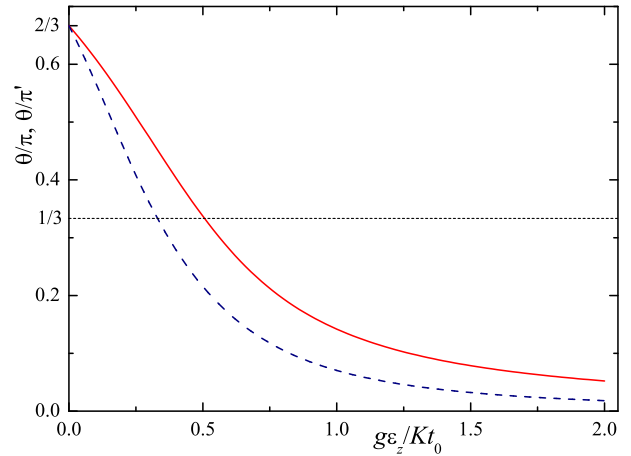}
\end{center}
\caption{\label{theta_bridge} (Color online) The dependence of $\theta_2$ (solid red curve) and $\theta'_2$ (dashed blue curve) on the stress $\varepsilon_z$.  The parameters of the model are: $g^2/Kt_0=1.2$, $K/g=40$.}
\end{figure}

First, let us discuss the behavior of orbital states at bridge sites. As it will be shown below (subsection D), the bridge sites can be characterized by the orbital state $|\theta\rangle$ corresponding to a definite angle $\theta$ in the orbital $(\tau^x, \tau^z)$ plane, whereas for the corner sites, the situation is more complicated. Let $\theta_2$ be the orbital angle for the bridge site 2. The corresponding local lattice distortions are given by an angle $\theta'_{2}$ in $(Q_2, Q_3)$ plane. The relationship between $\theta_{2}$ and $\theta'_{2}$ is given by Eq.~(\ref{Q}), which shows that $\theta'_{2} = \theta_{2}$ in the absence of external stress $\varepsilon_z$ but differ at $\varepsilon_z \neq 0$. In Fig.~\ref{theta_bridge}, we demonstrate the plots of $\theta_{2}$ and $\theta'_{2}$ as functions of $\varepsilon_z$. At $\varepsilon_z =0$, both angles are equal to $2\pi/3$ ($|2x^2-y^2-z^2\rangle$ orbital). With the growth of $\varepsilon_z$, $\theta'_{2}$ decreases faster than  $\theta_{2}$. At a certain value of $\varepsilon_z$, we have $\theta'_{2} = \pi/3$. Such $\theta'_{2}$ describes the oxygen octahedron compressed in $y$ direction and stretched along $x$ and $z$ axes. In the usual naive approach with $\theta'_{2} = \theta_{2}$, this should give $|x^2-z^2\rangle$ orbital lying in the ($x, z$) plane; this assumption was actually used in Ref.~\onlinecite{Huang} when the authors concluded that the orbital occupation in La$_{0.5}$Sr$_{1.5}$MnO$_4$ should be of $x^2-y^2/y^2-z^2$ -- type. However, in Fig.~\ref{theta_bridge}, we see that the  the actual orbital angle $\theta_{2}$ at $\theta'_{2} = \pi/3$  for the chosen values of parameters is about $\pi/2$, that is the electron density in $z$ direction is, in fact, much lower than for the $|x^2-z^2\rangle$ orbital. Physically this is due to the fact that the gain in kinetic energy related to the intersite electron hopping favors orbitals lying in the ($x, y$) plane, since the electron hopping occurs mainly in this plane.

\begin{figure}
\begin{center}
\includegraphics[width=0.95\columnwidth]{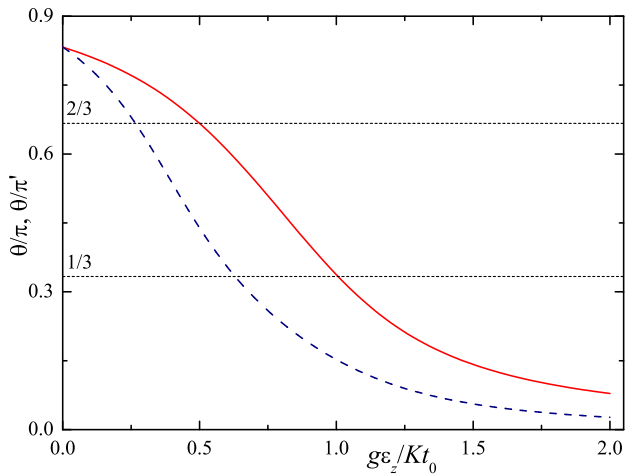}
\end{center}
\caption{\label{theta_Delta} (Color online) The dependence of $\theta_2$ (solid red curve) and $\theta'_2$ (dashed blue curve) on the stress $\varepsilon_z$ at non-zero crystal field splitting $\Delta = -0.5t_0$.  The parameters are the same as for Fig.~\ref{theta_bridge}.}
\end{figure}
\begin{figure}
\begin{center}
\includegraphics[width=0.95\columnwidth]{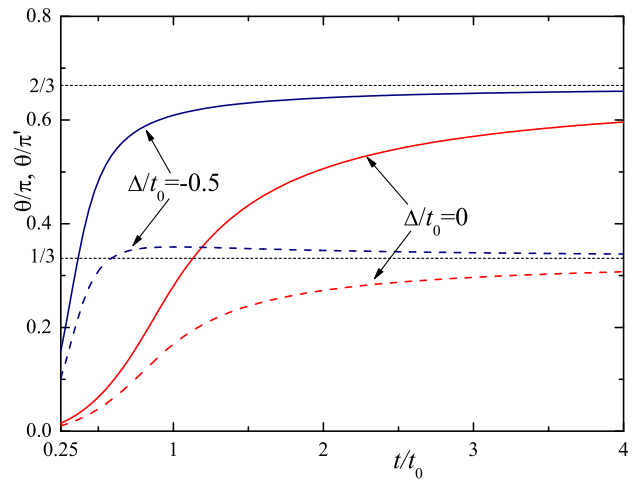}
\end{center}
\caption{\label{FigTheta_t} (Color online) The dependence of $\theta_2$ (solid curves) and $\theta'_2$ (dashed curves) on the hopping integral $t$ at zero and non-zero crystal field splitting $\Delta$. The stress $\varepsilon_z$ is chosen so that $g\varepsilon_z/Kt_0=0.6$. The parameters are the same as for Fig.~\ref{theta_bridge}.}
\end{figure}

Note that by including the term $\varepsilon_zQ_{3n}$ in Hamiltonian~\eqref{HCE}, we took into account a possible contribution to a crystal field splitting and, consequently, on the orbital occupation due to distortion of local ligand (here O$_6$ ) octahedron. At the same time, the crystal field  includes also a long-range interaction and it acts directly on the splitting of $e_g$ level. This additional splitting, which was introduced in Hamiltonian~\eqref{HCE} in the form $-\Delta\tau^z_n$, favors the orbitals lying in the ($x, y$) plane~\cite{HuaWu}. Indeed, as one can easily see, in the layered $214$ compounds like (LaSr)$_2$MnO$_4$ next nearest neighbors of a Mn ion will be four Mn$^{3.5+}$ in the basal plane (see Fig.~\ref{FigCE}), but no such ions (or lying at larger distances) in the $z$-direction. Consequently, the Coulomb field of these positively-charged ions in $ab$-plane would ``pull" the Mn electrons to this plane, i.e., they would cause extra crystal field splitting stabilizing ``flat" $x^2-y^2$ orbitals.

In Fig.~\ref{theta_Delta}, we demonstrate the effect of splitting $\Delta$ on the behavior of $\theta_{2}$ and $\theta'_{2}$ as functions of $\varepsilon_z$. We can see that for even relatively small values of $\Delta$, we have  $\theta_{2} \approx 2\pi/3$, i.e. the orbital occupation $|3y^2-r^2>$ (or $|3x^2-r^2>$) even at $\theta'_{2} = \pi/3$, i.e. for locally compressed (along $x$ or $y$ directions) MnO$_6$ octahedra. This agrees well with the experimental findings and {\it ab initio} calculation reported in Ref.~\onlinecite{HuaWu}, and explains its discrepancy with the results of Ref.~\onlinecite{Huang}.

As it was already mentioned above, for our layered system the kinetic energy gain determined by the in-plane electron hopping integral $t$ favors the orbitals lying in the ($x, y$) plane. In Fig.~\ref{FigTheta_t}, we show the dependence of $\theta_{2}$ and $\theta'_{2}$ on the hopping integral $t$. It is clearly seen that the difference between $\theta_{2}$ and $\theta'_{2}$ grows with $t$, and $\theta_{2}\to 2\pi/3$ whereas $\theta'_{2}$ still remains close to $\pi/3$ at large $t$.

\subsection{Lattice distortions}

Let us now analyze the JT distortions at the corner ($j=1,\,3$) and bridge ($j=2,\,4$) sites. We will consider the case of half-filling, $x=0.5$. The numerical calculations show that $\bar{Q}_{21}=0$. In the absence of the breathing mode, $Q_{1n}=0$, we can find the local distortions of Mn octahedron in $x$, $y$, and $z$ directions. They are related to $\bar{Q}_{2j}$ and $\bar{Q}_{3j}$ modes by the following formulas
\begin{eqnarray}
d^x_{j}&=&1+\frac{1}{\sqrt{2}}\left(\bar{Q}_{2j}
-\frac{\bar{Q}_{3j}}{\sqrt{3}}\right) \nonumber \\
d^y_{j}&=&1-\frac{1}{\sqrt{2}}\left(\bar{Q}_{2j}
+\frac{\bar{Q}_{3j}}{\sqrt{3}}\right) \nonumber \\
d^z_{j}&=&1+\sqrt{\frac{2}{3}}\bar{Q}_{3j}\,,\;\;j=1,\,2\,.
\end{eqnarray}

\begin{figure}
\begin{center}
\includegraphics*[width=0.95\columnwidth]{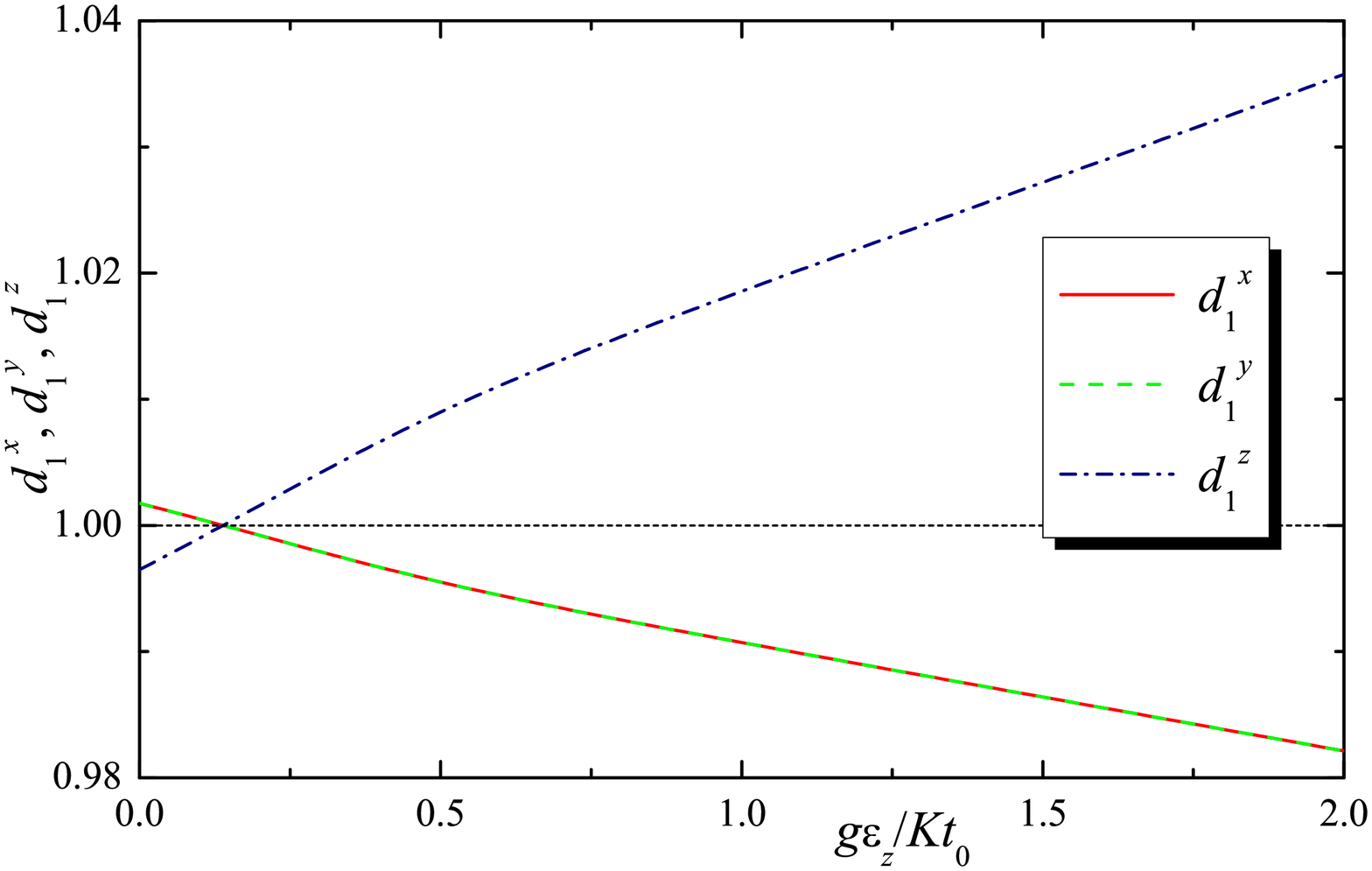}\\
\includegraphics*[width=0.95\columnwidth]{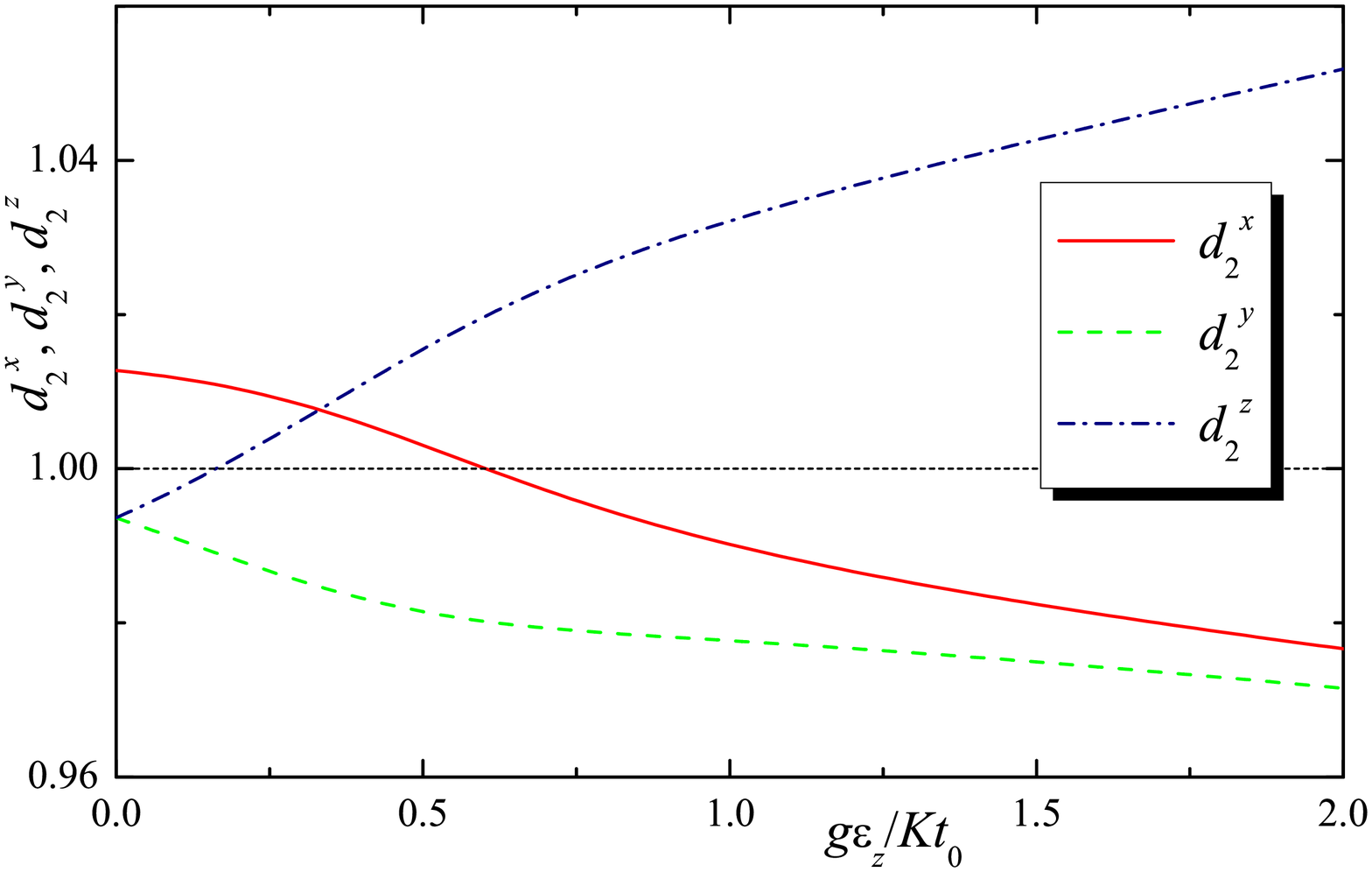}
\end{center}
\caption{\label{FigD} (Color online) The local distortions at the corner (upper panel) and bridge (lower panel) sites as a function of $\varepsilon_z$ (in units of $Kt_0/g$). The parameters are the same as for Fig.~\ref{theta_bridge}. For corner sites (upper panel) the distortions along $x$ and $y$ axes coincide.}
\end{figure}

The values of $d^{x,y,z}_j=1$ correspond to an undistorted octahedron. For $j=3,\,4$, one needs to use relationships~\eqref{Qsymm}. The dependence of the local distortions on the stress $\varepsilon_z$ is shown in Fig.~\ref{FigD}. At $\varepsilon_z=0$, for the corner site $1$ we have compressed octahedron along $z$ axis and stretched equally along $x$ and $y$ axes. With the growth of the stress, the deformation (elongation) in $z$ direction increases, and at some critical value of $\varepsilon_z$ the deformations $x$, $y$ and $z$ change sign, and the octahedron around corner site $1$ becomes elongated.

For the bridge site $2$, we have compression in $y$ and $z$ direction, and stretching in $x$ direction at small $\varepsilon_z$. With increasing $\varepsilon_z$. the elongation in $z$ direction increases, and deformations in $x$ and $y$ directions decrease. Eventually, for very large $\varepsilon_z$, we would have stretched octahedra along the $z$ axis. But for intermediate values of $\varepsilon_z$. we may have the situation with the distances (MnO)$_z$ = (MnO)$_y$ $>$ (MnO)$_x$ (or, for the other edge site, (MnO)$_z$ = (MnO)$_x$ $>$ (MnO)$_y$), i.e. the locally-compressed (along $x$ or $y$ directions) MnO$_6$ octahedra. As we saw above, in Fig.~\ref{FigTheta_t}, the form of occupied orbitals in this case may still be ($3y^2-r^2$) (or ($3x^2-r^2$)). Experimentally this indeed seems to be the case in La$_{0.4}$Sr$_{1.5}$MnO$_4$~\cite{HuaWu}.

\subsection{Charge disproportionalization}

\begin{figure}
\begin{center}
\includegraphics*[width=0.95\columnwidth]{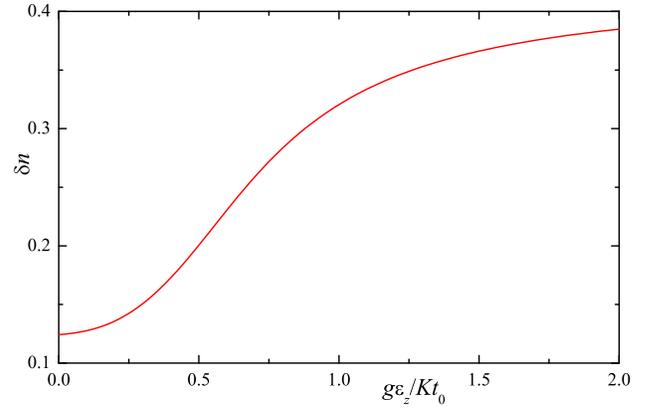}
\end{center}
\caption{\label{FigDeltaN} (Color online) The disproportionalization $\delta n$ in electron density between the bridge and corner sites vs. $\varepsilon_z$. The parameters are the same as for Fig.~\ref{theta_bridge}.}
\end{figure}

Let us now consider the electronic degrees of freedom. Note first of all that there exists a charge disproportionalization between corner and bridge sites. To measure this disproportionalization, we introduce the variable $\delta n=\langle c^{\dag}_{n2a}c_{n2a}\rangle+\langle c^{\dag}_{n2b}c_{n2b}\rangle-n$ which describes the deviation in the electron density at the bridge site from mean value $n$. The dependence of $\delta n$ on $\varepsilon_z$ at $n=1-x=0.5$ is shown in Fig.~\ref{FigDeltaN}. We see, that $\delta n$ is always positive, that is, the electrons more likely occupy bridge sites. We also see that the charge-transfer $\delta n$ is always less that 1, even for the (unphysically) large strain, {\it cf} Ref.~\onlinecite{Yarlag}.

\subsection{Occupation of orbital states}

Let us now analyze in more detail the orbital configurations at the corner and bridge sites. First of all, until now we considered ``pure'' orbital states, assuming that the electrons occupy a particular orbital. However, strictly speaking, for non-zero hopping the on-site orbital state can not be described by the orbital wave functions $|\theta\rangle$ of the form of Eq.~\eqref{theta}: band formation can lead to mixing of orbital occupation, and, strictly speaking, it is necessary to consider a density matrix $\hat{\rho}^{j}$ ($j=1,\,2,\,3,\,4$). In the basis of operators $a_{n_j\alpha}$, it can be written as
\begin{equation}\label{rho}
\hat{\rho}^j_{\alpha\beta}=\langle a_{n_j\beta}^{\dag}a_{n_j\alpha}\rangle\,,
\end{equation}
where $n_j$ corresponds to the corner site if $j=1,\,3$, or to the bridge site when $j=2,\,4$. The elements of this matrix can be calculated using relationships between $a_{n_j\alpha}$ and $c_{nj\alpha}$ operators, and Eq.~\eqref{meanC} for $\langle c^{\dag}_{nA}c_{nB}\rangle$. The JT distortions $Q_{2n}$ and $Q_{3n}$ can be expressed in terms of the density matrix. On the other hand the given $Q_{2n}$ and $Q_{3n}$ can not provide unambiguous determination of orbital states, since the orbital states themselves are mixed.

In the general case, the density matrix has non-zero complex conjugate non-diagonal elements. It can be diagonalized using orthogonal matrix of the form
\begin{equation}\label{Rtheta}
\hat{R}_{\theta}=\left(%
\begin{array}{cc}
\cos\frac{\theta}{2} & \sin\frac{\theta}{2} \\
-\sin\frac{\theta}{2} & \cos\frac{\theta}{2} \\
\end{array}%
\right),
\end{equation}
and~\footnote{The inequality $n^{\theta}_{ja}>n^{\theta}_{jb}$ can be always satisfied. Indeed, if the matrix $\hat{R}_{\theta}$ diagonalizes $\hat{\rho}^{j}$ with $n^{\theta}_{ja}<n^{\theta}_{jb}$, then the matrix $\hat{R}_{\theta+\pi}$ also diagonalizes $\hat{\rho}^{j}$ with $n^{\theta+\pi}_{ja}=n^{\theta}_{jb}>
n^{\theta+\pi}_{jb}=n^{\theta}_{ja}$.}
\begin{equation}\label{rhotheta}
\hat{R}_{\theta}^{\dag}\hat{\rho}^j\hat{R}_{\theta}=\left(%
\begin{array}{cc}
n^{\theta}_{ja} & 0 \\
0 & n^{\theta}_{ja} \\
\end{array}%
\right),\;\;n^{\theta}_{ja}>n^{\theta}_{jb}\,.
\end{equation}

\begin{figure}
\begin{center}
\includegraphics*[width=0.95\columnwidth]{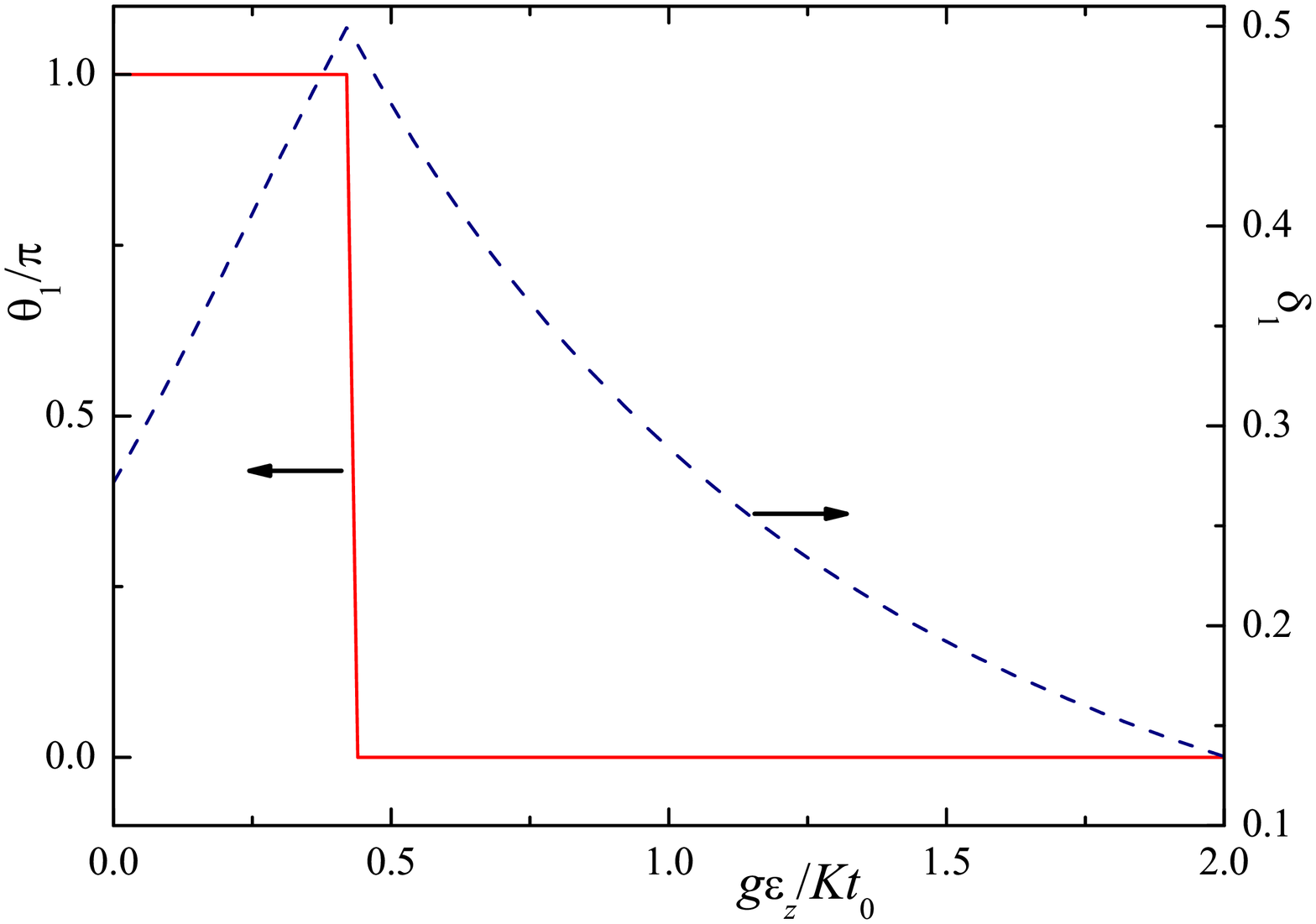}\\
\includegraphics*[width=0.95\columnwidth]{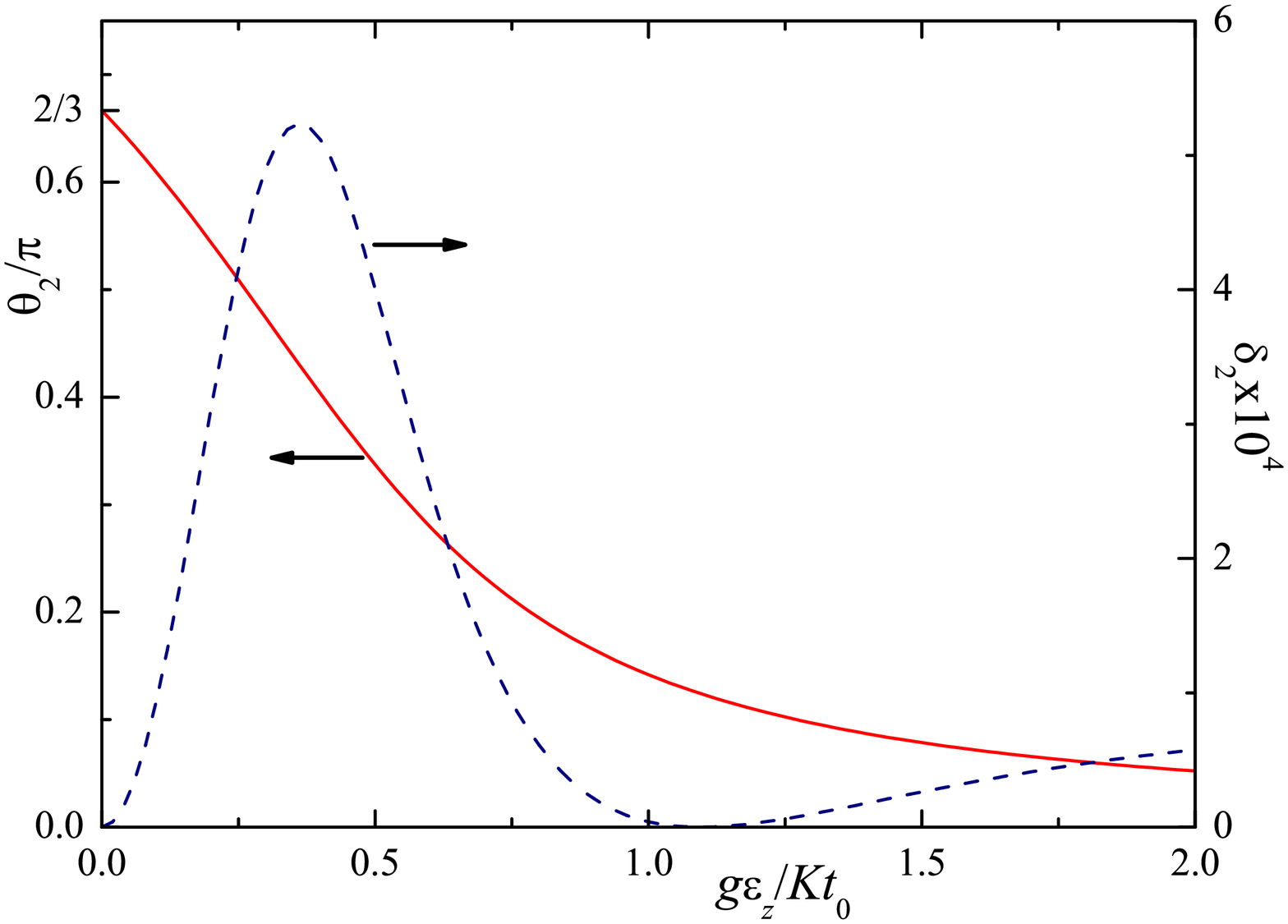}
\end{center}
\caption{\label{FigTheta} (Color online) The dependence of $\theta_j$ (red curves) and $\delta_j$ (blue curves) on the stress $\varepsilon_z$ calculated for the corner ($j=1$, upper panel) and the bridge ($j=2$, lower panel) sites. The parameters are the same as for Fig.~\ref{theta_bridge}.}
\end{figure}

Transformation~\eqref{rhotheta} corresponds to rotation in the orbital space by an angle $\theta$. It is easy to show that the value $n^{\theta}_{ja}$ corresponds to the maximum possible diagonal element of the set of matrices of the form $\hat{R}_{\theta}^{\dag}\hat{\rho}^j\hat{R}_{\theta}$, that is, the orbital state $|\theta_j\rangle$ is mostly occupied.  But since $n^{\theta_j}_{jb}\neq0$, the orthogonal state $|\theta_j+\pi\rangle$ is also occupied. The probabilities $p_{ja}$ and $p_{jb}$ to find an electron at a site $n_j$ in states $|\theta_j\rangle$ and $|\theta_j+\pi\rangle$, respectively, are the following
\begin{equation}\label{p}
p_{ja,b}=\frac12\pm\frac{\sqrt{\langle{\tau}^{x}_{nj}\rangle^2
+\langle{\tau}^{z}_{nj}\rangle^2}}%
{2\sum\limits_{\alpha}n_{j\alpha}}\,,
\end{equation}
where plus (minus) sign corresponds to the $p_{ja}$ ($p_{jb}$). We will consider $\delta_j\equiv p_{jb}$ as a measurements of deviation of the electronic state at a site $n_j$ from the pure state $|\theta_j\rangle$. In Fig.~\ref{FigTheta} the dependence of $\theta_j$ and $\delta_j$ on the stress $\varepsilon_z$ is shown both for the corner ($j=1$, upper panel) and the bridge ($j=2$, lower panel) sites. The angle $\theta_1$ for the corner site changes suddenly from the value $\theta_1=0$ to $\theta_1=\pi$ at some critical value of the stress $\varepsilon_z$, while the $\theta_2$ for the bridge site change continuously from $\theta_2=2\pi/3$ to $\theta_2=0$.

We see that for the bridge site the state $|\theta_2+\pi\rangle$ is almost empty ($\delta_2<10^{-3}\ll1$) for any value of $\varepsilon_z$, while for the corner sites the probabilities $p_{1a}$ and $p_{1b}$ turn out to be of the same order of magnitude. That is, for the bridge sites, we have nearly pure orbital states $|\theta_{2,4}\rangle$, thus justifying our previous treatment, in which we predominantly considered orbital occupation and distortion of the bridge sites. At the same time, for the corner sites both orbitals are populated. This corresponds to the conclusions reached in Ref.~\onlinecite{Yarlag}.

\subsection{Discussion of the results}

In our calculations, we modeled the real situation met in layered manganites La$_{2-x}$Sr$_x$MnO$_4$, especially for the composition $x=1.5$, for which the system exibits charge, orbital, and magnetic ordering of the CE type. We have shown that the orbital occupation of the bridge sites ``Mn$^{3+}$" can be treated as an occupation of a particular ``pure" orbital state, characterized by the mixing angle $\theta$, Eq.~\eqref{theta}, whereas band effects make orbital occupation of corner sites less well defined: both orbitals have comparable occupation at corner sites. This follows just from the geometry of the system, with its ferromagnetic zigzags: as argued in Ref.~\onlinecite{vdBrKhaKho}, for bridge sites electrons from only one orbital can hop to its corner neighbors, whereas at corner sites, both orbitals, e.g. $|x^2-y^2\rangle$ and $|z^2\rangle$  participate in the band formation. We have also shown that the degree of charge transfer in the charge-ordered state is (much) less than 1, typically $\delta n$ is about 0.2.

However, the most important is the conclusion that in a general case the type of occupied orbitals may strongly differ from what one would deduce from the local distortion of MO$_6$ octahedra. That is, the usually assumed one-to-one correspondence between the orbital occupation and the local JT distortion, very often used to determine orbital occupation from structural data, may in general break down.

Two main factors can lead to this effect. One is the contribution of further neighbors to a crystal field splitting. Thus, as explained above, in layered materials like La$_{2-x}$Sr$_x$MnO$_4$ there are more positively-charged Mn$^{3+}$/Mn$^{4+}$ ions in $xy$ plane than in $z$ direction, which would move the $x^2-y^2$ level down, even though local MnO$_6$ octahedra may be somewhat elongated in $z$ direction, e.g. due to strain, which is always present in layered systems. Actually, similar effect was already noticed for some other systems, e.g. for BaCoO$_3$~\cite{Pardo} or for LaTiO$_3$~\cite{Pavarini}.

Another very important factor is the role of electron kinetic energy. For layered systems one would gain maximum kinetic energy if the relevant orbitals are ``put in plain'' - again despite the fact that local octahedra may be elongated in perpendicular direction. This factor is more important for larger hopping, notably in systems approaching localized-itinerant crossover (Mott transition). Apparently the considered system La$_{2-x}$Sr$_x$MnO$_4$ is in this regime.

In effect, both these factors lead to the situation, in which the orbital occupation may strongly differ from that which one would deduce from the structural data. Apparently this is what happens in La$_{0.5}$Sr$_{1.5}$MnO$_4$, in which the MnO$_6$ octahedra are distorted so that the Mn-O distance to the apex oxygen is large (practically equal to the long Mn-O distance in plane). This distortion would in the usual picture lead to the occupation of $x^2-z^2$ or $y^2-z^2$ orbitals, which was indeed proposed in Ref.~\onlinecite{Huang}. However due to the factors discussed above the actual orbital occupation may be quite different, and indeed it was observed experimentally and confirmed by {\it ab initio} band structure calculations that the occupied orbitals in this case are rather of $2x^2-y^2-z^2$ or $2y^2-x^2-z^2$ type~\cite{HuaWu}. Our model calculations show that it is indeed possible due to these two factors mentioned above. Thus, our results demonstrate that in general the orbital occupation may strongly differ from that expected from local JT distortions - especially in anisotropic systems and systems close to the localized-itinerant crossover. Thus, one has to be very careful in using the conventional method to determine orbital structure from lattice distortions: in some cases this often applied method can give quite wrong results.

\section{Conclusion}

Through the analysis of the lattice distortions and orbital structure of half-doped manganites, we demonstrated that in systems with orbital degeneracy and orbital ordering the type of orbital occupation may strongly deviate from the one which would be deduced from the local JT distortion, i.e. the conventional JT physics may be violated in some cases. Thus, one has to be very careful in using the standard, widely used method of determining orbital occupation from the corresponding JT distortion. To obtain correct results, it is necessary to consider self-consistently orbital occupation, lattice distortions, and charge disproportionalization.

\section*{Acknowledgments}

The work was supported by the Russian Foundation for Basic Research (projects 07-02-91567 and 08-02-00212), and by the Deutsche Forschungsgemeinshaft via SFB 608 and the German-Russian project 436 RUS 113/942/0.

\end{document}